# TRIOT: Faster tensor manipulation in C++11

Florian Heyl[a] and Oliver Serang[a]

a   Informatik u. Mathematik, Freie Universität Berlin, Germany

**Abstract**   **Context:** Multidimensional arrays are used by many different algorithms. As such, indexing and broadcasting complex operations over multidimensional arrays are ubiquitous tasks and can be performance limiting.

**Inquiry:** Simultaneously indexing two or more multidimensional arrays with different shapes (*e.g.*, copying data from one tensor to another larger, zero padded tensor in anticipation of a convolution) is difficult to do efficiently: Hard-coded nested for loops in C, Fortran, and Go cannot be applied when the dimension of a tensor is unknown at compile time. Likewise, boost::multi_array cannot be used unless the dimensions of the array are known at compile time, and the style of implementation restricts the user from using the index tuple inside a vectorized operation (as would be required to compute an expected value of a multidimensional distribution). On the other hand, iteration methods that do not require the dimensionality or shape to be known at compile time (*e.g.*, incrementing and applying carry operations to index tuples or remapping integer indices in the flat array), can be substantially slower than hard-coded nested for loops.

**Approach:** Using a few tasks that broadcast operations over multidimensional arrays, several existing methods are compared: hard-coded nested for loops in C and Go, vectorized operations in Fortran, boost::multi_array, numpy, tuple incrementing, and integer reindexing.

**Knowledge:** A new approach to this problem, "template-recursive iteration over tensors" (TRIOT), is proposed. This new method, which is made possible by features of C++11, can be used when the dimensions of the tensors are unknown at compile time. Furthermore, the proposed method can access the index tuple inside the vectorized operations, permitting much more flexible operations.

**Grounding:** Runtimes of all methods are compared, and demonstrate that the proposed TRIOT method is competitive with both hard-coded nested for loops in C and Go, as well as vectorized operations in Fortran, despite not knowing the dimensions at compile time. TRIOT outperforms boost::multi_array, numpy, tuple incrementing, and integer reindexing.

**Importance:** Manipulation of multidimensional arrays is a common task in software, especially in high-performance numerical methods. This paper proposes a novel way to leverage template recursion to iterate over and apply operations to multidimensional arrays, and then demonstrates the superior performance and flexibility of operations that can be achieved using this new approach.

## The Art, Science, and Engineering of Programming

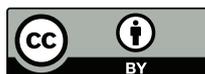

**TRIOT: Faster tensor manipulation in C++11**

## 1 Introduction

Tensors (*i.e.*, multidimensional arrays) are extremely important in many areas of research, including physics [4] (where tensors prominently feature in numerical simulations in a discretized space), statistics [3] (where tensors represent multidimensional probability mass functions and where tensor convolutions are bijective to sums and differences between random variables), and signal processing [1] (where tensors are used to represent multidimensional images, *e.g.*, tomography on magnetic resonance imaging scans). The most fundamental tasks for tensor code are to efficiently retrieve data from a particular index and to use those indexing routines to vectorize many such indexing operations (and possible modifications at each index).

However, where 1-dimensional vector lookup is simple (it can be performed as a pointer addition and a pointer dereference or as a single LDR load relative assembly instruction), tensor indexing is more complicated, because indices are integer tuples (note: this manuscript uses "tuple" in the mathematical sense rather than as a type-heterogeneous list as denoted by std::tuple). For example accessing element $(2, 0, 4, 1)$ of a tensor with the shape $(4, 9, 7, 5)$ would be performed by first computing that element's linear memory offset (the integral "flat index"), $2 \cdot 9 \cdot 7 \cdot 5 + 0 \cdot 7 \cdot 5 + 4 \cdot 5 + 1$. Although converting a tuple index to a flat integer index can be avoided when vectorizing operations over the flat array (*e.g.*, performing an element-wise multiplication on every element in a tensor can be done by simply iterating over its flat integer indices), the flat integer indices are not sufficient when vectorizing operations over two or more tensors of different shapes (this is very common when multiplying two probability mass functions with different support or when zero padding a tensor to perform convolution via the multidimensional fast Fourier transform). In that case, tuple indexing must be performed. Consider a tensor $x$ with dimension $d$ and shape $x.shape = (s_0, s_1, \ldots, s_{d-1})$. Lookup of index tuple $t = (t_0, t_1, \ldots, t_{d-1})$ (where $\forall i \in \{0, 1, \ldots, d-1\}, \ t_i \in \{0, 1, \ldots, s_i - 1\}$) is performed by computing the corresponding integer index in the flat vector $x.flat$:

$$\begin{aligned} x[t] = x.flat\big[ & t_0 \cdot s_1 \cdot s_2 \cdot s_3 \cdot s_4 \cdots s_{d-1} + \\ & t_1 \cdot s_2 \cdot s_3 \cdot s_4 \cdots s_{d-1} + \\ & t_2 \cdot s_3 \cdot s_4 \cdots s_{d-1} + \cdots \\ & \cdots + t_{d-2} \cdot s_{d-1} \\ & + t_{d-1}\big], \end{aligned}$$

assuming row-major order of the flat vector (as used in the C programming language). This can be rewritten more efficiently using Horner's rule:

$$x[t] = x.flat[(\cdots(((t_0 \cdot s_1 + t_1) \cdot s_2 + t_2) \cdot s_3 + t_3) \cdot s_4 + \cdots + t_{d-2} \cdot s_{d-1}) + t_{d-1}].$$

The classic manner for iterating over the tensor $x$ is therefore to use nested for loops (**Algorithm 1**).

However, to write those nested for loops, it is necessary at compile time to know both the number of nested loops (which means it is necessary to know the dimensionality

■ **Algorithm 1** Iterating over a tensor with nested for loops. The algorithm iterates over tensor *x* with shape *x.shape*, where *x.dimension* is the dimensionality of *x*.

```c
typedef unsigned long*__restrict const tup_t;
typedef const unsigned long*__restrict const const_tup_t;

void nested_for_loops(const_tup_t shape) {
  // Writing x.dimension separate nested for loops:
  for (unsigned long i0=0; i0<shape[0]; ++i0)
    for (unsigned long i1=0; i1<shape[1]; ++i1)
      for (unsigned long i2=0; i2<shape[2]; ++i2)
        // ...
        {
          // Inside innermost loop:
          unsigned long x_index = ((i0*shape[1] + i1)*shape[2] + i2)*shape[3] /* + ...
                         using each loop variable once */ ;

          // Perform operations on x.flat[x_index] for some tensor x in global scope:
        }
}
```

■ **Algorithm 2** Tuple to flat index with respect to shape. The algorithm takes a tuple tup and a shape shape, both of length dimension, and returns the flat index of tup with respect to shape.

```c
unsigned long tuple_to_index(const_tup_t tup, const_tup_t shape, unsigned int dimension) {
  unsigned long res = 0;
  unsigned int k;
  for (k=0; k<dimension−1; ++k) {
    res += tup[k];
    res *= shape[k+1];
  }
  res += tup[k];
  return res;
}
```

*d*) and the operations that will be performed inside the loops. Of course, if a method is simply applying a single operation to each index (*e.g.*, x[x_index] += 1), the loop can be performed by iterating with an integer index over the flat index values $\in \{0, 1, 2, \ldots, s_0 \cdot s_1 \cdot s_2 \cdots s_{d-1} - 1\}$. But when performing operations on tensors of different shapes, that is not a viable solution. For instance, $\forall$ valid tuples *t* in range of shape *s*, x[t] += y[t] (or equivalently x[x_index] += y[y_index] for the appropriate flat index values) could be performed by initializing x_index=0 and incrementing it inside the innermost for loop. This would allow x_index to be computed trivially, but y_index would still need to be computed as x_index is in **Algorithm 1**.

When the dimensionality *d* is not known at compile time, one can consider a tuple index stored in an array. Looking up the flat index value can be performed in a simple C-style routine ( **Algorithm 2**).

What is less trivial is advancing the tuple index to the next lexicographic tuple index (note that other advancing strategies may be possible, but lexicographic ordering

■ **Algorithm 3** Advance tuple with respect to shape. The algorithm takes a tuple tup and a shape shape, both of length dimension, and advances tup to its next lexicographic value with respect to shape.

```
1  void advance_tuple(tup_t tup, const_tup_t shape, unsigned int dimension) {
2    ++tup[dimension−1];
3
4    for (unsigned int k=dimension−1; k>=1; −−k)
5      if ( tup[k] >= shape[k] ) {
6        // Perform carry operation:
7        ++tup[k−1];
8        tup[k] = 0;
9      }
10     else
11       // No more carry operations:
12       return;
13 }
```

corresponds to the ideal order for caching). The lexicographic ordering can be iterated through just as one would count numbers with carry operations (**Algorithm 3**).

For a problem of dimension $d$, this advance_tuple routine runs in $O(d)$ in the worst case, but in the average case it is more efficient because it returns early once there are no further carry operations to perform. Consider that for a random valid tuple $t$ with respect to shape $s$, the chances of returning after 1 iteration (*i.e.*, performing no carry operations at all) will be $\frac{s_{d-1}-1}{s_{d-1}}$, because that carry operation will only be performed when $t_{d-1} = s_{d-1}$ before calling advance_tuple. The chances of performing the second carry are even smaller $\frac{s_{d-1}-1}{s_{d-1}} \cdot \frac{s_{d-2}-1}{s_{d-2}}$, so that the chances of progressing to each next carry will decay exponentially for tensors with non-trivial shapes (*i.e.* for tensors with no axis $i$ along which $s_i = 1$); Therefore, the average number of iterations required by the for loop in **Algorithm 3** will be a constant (because it is bounded above by summing a geometric series). Since each valid tuple will be visited, this average cost will equal the amortized cost, and the runtime of advance_tuple will be in (1).

However, at least one if statement is needed for each call to advance_tuple, and so the runtime can be seriously affected by the extra BR branch line in the resulting assembly code. Because of this and because the loop in advance_tuple cannot be easily unrolled, using tuple iteration and the advance_tuple routine can be significantly slower than performing the equivalent nested for loops.

An alternative to tuple iteration is to use a number-theoretic strategy, which iterates over a single flat index with respect to a first shape and then preforms reindexing to convert it to a flat index with respect to another shape. For some unknown tuple $t$, suppose x_index is the flat integer index with respect to shape $s$ and y_index is the flat integer index with respect to shape $s'$:

$$\begin{aligned}\text{x\_index} = &\, t_0 \cdot s_1 \cdot s_2 \cdot s_3 \cdot s_4 \cdots s_{d-1} + \\ & t_1 \cdot s_2 \cdot s_3 \cdot s_4 \cdots s_{d-1} + \\ & t_2 \cdot s_3 \cdot s_4 \cdots s_{d-1} + \cdots \\ & \cdots + t_{d-2} \cdot s_{d-1} \\ & + t_{d-1}\end{aligned}$$

Consider that in the sum to compute x_index, every term is multiplied with $s_{d-1}$ except $t_{d-1}$, meaning that x_index $\equiv t_{d-1} \pmod{s_{d-1}}$. Furthermore, for every axis $i$, a valid tuple $t$ with respect to shape $s$ is always in range $\{0, 1, 2, \ldots, s_{d-1} - 1\}$. Together these imply that $t_{d-1} = $ x_index $\% \, s_{d-1}$. Subtracting the computed value of $t_{d-1}$ from x_index and dividing by $s_{d-1}$ yields an expression of the same form, and so it is possible to compute the full tuple $t$. From there, it is possible to compute the value of y_index:

$$\begin{aligned}\text{y\_index} = &\, t_0 \cdot s'_1 \cdot s'_2 \cdot s'_3 \cdot s'_4 \cdots s'_{d-1} + \\ & t_1 \cdot s'_2 \cdot s'_3 \cdot s'_4 \cdots s'_{d-1} + \\ & t_2 \cdot s'_3 \cdot s'_4 \cdots s'_{d-1} + \\ & \vdots \\ & + t_{d-2} \cdot s'_{d-1} \\ & + t_{d-1}.\end{aligned}$$

These steps can be combined into a single loop that computes y_index while it computes each next $t_i$ ( **Algorithm** 4). This type of reindexing means that for some applications it is possible to completely avoid storing the tuple and simply iterate over flat integer indices in one tensor and then map each to its appropriate flat integer index in the other tensor. Although this method seems promising, it simply moves the computational cost from advance_tuple into reindex, and because division and modulo operation are almost always slower than multiplication and addition, this method is slower than tuple iteration in practice.

Here a novel pattern is proposed for efficiently vectorizing over tensors in C++11. This strategy allows iterating over tuples with speed comparable to the C-style nested for loops but with the generality of tuple iteration and where the dimensionality does not need to be known at compile time. The proposed method does require that the maximum tensor dimension to be fixed at compile time (as with languages like Fortran); however, as we will show, this is not limiting in practice, as the memory usage is superexponential in the dimension. The proposed method is fairly involved to derive, but can be used in an intuitive, flexible manner via the provided header-only library.

■ **Algorithm 4** Reindex flat index from one tensor shape to another tensor shape. The algorithm takes a flat integer index with respect to shape and converts it to a flat integer index with respect to new_shape, where both shapes have length dimension.

```
1  inline unsigned long reindex(unsigned long index, const_tup_t shape, const_tup_t new_shape, unsigned
       ↪ int dimension) {
2    unsigned long new_index = 0;
3    unsigned long new_axis_product_from_right = 1;
4    for (int i=dimension—1; index>0 && i>=0; ——i) {
5      unsigned long next_axis = shape[i];
6      unsigned long new_next_axis = new_shape[i];
7
8      unsigned long next_value = index % next_axis;
9
10     new_index += next_value * new_axis_product_from_right;
11     index /= next_axis;
12
13     new_axis_product_from_right *= new_next_axis;
14   }
15   return new_index;
16 }
```

## 2 Methods

It is first assumed the dimension of the tensors, denoted DIMENSION, is known at compile time (*i.e.*, it is constexpr). Later, that implementation is generalized to allow the dimension to be known only at runtime. First, **Algorithm 2** is re-implemented to exploit compile-time knowledge of the dimension (**Algorithm 5**). Second, nested for loops over a given shape are produced via template recursion ( **Algorithm 6**). Note that this proposed strategy is distinct from fully unrolling loops on flat integer indices via template recursion; while forced loop unrolling can sometimes be fast, it can also lead to an enormous executable (which in turn can also burden the cache and end up making slower code).

When invoking ForEachFixedDimension<DIMENSION>::template apply, the typename FUNCTION and the template parameter pack typename ...TENSORS can be inferred at compile time based on the contents of the tensor args... and the argument types to function. Also note that function can have any valid callable type: C-style function pointer, std::function, or any functor (*i.e.*, any object that defines an () operator). This so-called "duck typing", means that the method can be used whenever valid type substitutions can be found. It also offers variadic support while ensuring type safety (unlike the unsafe C-style ... used to implement a varargs function like printf). Note that the tuple value can be safely declared __restrict, meaning that it is referred to from no other location in memory, because it will be constructed specifically for iteration and then deallocated.

Template recursion can likewise be used to convert a runtime value into a compile-time constant (**Algorithm 7**). Note that here, even though template recursion is used, the dimension need not be known until runtime. This is essentially achieved by using templates as a form of just-in-time (JIT) compilation, precomputing strategies for all

■ **Algorithm 5** Tuple to flat index with respect to shape (special case for fixed dimension). Equivalent to **Algorithm 2**, but where the dimension is restricted to a constant known at compile time. On some newer compilers, this will not offer a substantial speedup, because aggressive constant propagation will already recognize cases where DIMENSION is known at compile time.

```
1  template <unsigned int DIMENSION>
2  inline unsigned long tuple_to_index_fixed_dimension(const_tup_t tup, const_tup_t shape) {
3    unsigned long res = 0;
4    unsigned int k;
5    for (k=0; k<DIMENSION−1; ++k) {
6      res += tup[k];
7      res *= shape[k+1];
8    }
9    res += tup[k];
10   return res;
11 }
```

■ **Algorithm 6** Nesting loops via template recursion. Nested for loops are unrolled while keeping the tuple iterated over in a contiguous array for later use.

```
1  template <unsigned int DIMENSION, unsigned int CURRENT>
2  class ForEachFixedDimensionHelper {
3  public:
4    template <typename FUNCTION, typename ...TENSORS>
5    inline static void apply(tup_t counter, const_tup_t shape, FUNCTION function, TENSORS & ...args) {
6      for (counter[CURRENT]=0; counter[CURRENT]<shape[CURRENT]; ++counter[CURRENT])
7        ForEachFixedDimensionHelper<DIMENSION−1, CURRENT+1>::template apply<FUNCTION, TENSORS...>(
           ↪ counter, shape, function, args...);
8    }
9  };
10
11 template <unsigned int CURRENT>
12 class ForEachFixedDimensionHelper<1u, CURRENT> {
13 public:
14   template <typename FUNCTION, typename ...TENSORS>
15   inline static void apply(tup_t counter, const_tup_t shape, FUNCTION function, TENSORS & ...args) {
16     for (counter[CURRENT]=0; counter[CURRENT]<shape[CURRENT]; ++counter[CURRENT])
17       function(args[tuple_to_index_fixed_dimension<CURRENT+1>(counter, args.data_shape())]...);
18       /* tensor.data_shape() is an accessor for returning the shape member. */
19   }
20 };
21
22 template <unsigned char DIMENSION>
23 class ForEachFixedDimension {
24 public:
25   template <typename FUNCTION, typename ...TENSORS>
26   inline static void apply(const_tup_t shape, FUNCTION function, TENSORS & ...args) {
27       unsigned long counter[DIMENSION];
28       memset(counter, 0, DIMENSION*sizeof(unsigned long));
29       ForEachFixedDimensionHelper<DIMENSION,0>::template apply<FUNCTION, TENSORS...>(counter,
           ↪ shape, function, args...);
30   }
31 };
```

■ **Algorithm 7** Dynamic linear template lookup. Searches linearly from MINIMUM to MAXIMUM to find a compile-time constant matching a value known only at runtime. This essentially converts runtime integer values into constexpr values.

```
1  typedef unsigned int TEMPLATE_SEARCH_INT_TYPE;
2  template <TEMPLATE_SEARCH_INT_TYPE MINIMUM, TEMPLATE_SEARCH_INT_TYPE MAXIMUM, template <
   ↪ TEMPLATE_SEARCH_INT_TYPE> class WORKER>
3  class LinearTemplateSearch {
4  public:
5    template <typename...ARG_TYPES>
6    inline static void apply(TEMPLATE_SEARCH_INT_TYPE v, ARG_TYPES && ... args) {
7      if (v == MINIMUM)
8        WORKER<MINIMUM>::apply(std::forward<ARG_TYPES>(args)...);
9      else
10       LinearTemplateSearch<MINIMUM+1, MAXIMUM, WORKER>::apply(v, std::forward<ARG_TYPES>(args)...);
11   }
12 };
13
14 template <TEMPLATE_SEARCH_INT_TYPE MAXIMUM, template <TEMPLATE_SEARCH_INT_TYPE> class WORKER
   ↪ >
15 class LinearTemplateSearch<MAXIMUM, MAXIMUM, WORKER> {
16 public:
17   template <typename...ARG_TYPES>
18   inline static void apply(TEMPLATE_SEARCH_INT_TYPE v, ARG_TYPES && ... args) {
19     assert(v == MAXIMUM);
20     WORKER<MAXIMUM>::apply(std::forward<ARG_TYPES>(args)...);
21   }
22 };
```

dimensionalities of interest and then looking up the correct one at runtime. As shown, this runtime lookup will be unrolled into a series of branches that will give a lookup runtime linear in the dimension queried. Note that log search (rather than linear search) would also be possible, but it would not amortize out the runtime the same: here only $O(d)$ steps are used, which is already a prerequisite cost for processing $d$-dimensional tensors. Likewise, instead of linear search, virtual functions can be used via a base class that offers an interface for shared functionality regardless of $d$. This can be used to construct a table of derived objects (or a table of factories) where the compile-time constant corresponds to its index in the table, and then the table can be queried to retrieve a pointer to a derived object (cast as base class pointer). Although virtual functions would achieve this lookup in $O(1)$ rather than $O(d)$, they shield function calls from inlining and thus can impede compiler optimizations. For these reasons, the linear search is preferred.

**Algorithm 6** and **Algorithm 7** can be invoked together to efficiently iterate over a tuple with dimension known only at runtime (**Algorithm 8**). This can be called elegantly with $\lambda$-functions from C++11, which can even modify external state using variable capture. The use of $\lambda$-functions and variadic templates (which allow expressions to be mapped over an arbitrary number of tensors) are essential, and are the reasons that this pattern is restricted to C++11.

Using the same approach, similar functions can be defined to allow modification (or modification of only one argument as implemented in apply_tensors) of tensor data by

■ **Algorithm 8** Template-recursive iteration over tensors (TRIOT). Pairing **Algorithm 6** with **Algorithm 7** enables mapping of functions over tensors. An example is shown that uses variable capture to compute the inner product between two tensors of different shape. The maximal tensor dimension of 32 will be discussed later on.

```
/* check_tensor_pack_bounds scrutinizes if the dimensions match between the two tensors and controls
    ↪ that the shape of the second tensor is in bounds with the first tensor. */

const unsigned int MAX_TENSOR_DIMENSION = 32u;
template <typename FUNCTION, typename ...TENSORS>
/* Note: using simple in-house vector type (not std::vector), which is also used as tensor.flat */
template <typename FUNCTION, typename ...TENSORS>
void for_each_tensors(FUNCTION function, const Vector<unsigned long> & shape, const TENSORS & ...args)
    ↪ {
  check_tensor_pack_bounds<TENSORS...>(args..., shape);
  LinearTemplateSearch<0u,MAX_TENSOR_DIMENSION,TRIOT::ForEachFixedDimension>::apply(shape.size(),
      ↪ shape, function, args...);
}

double dot_product(const tensor & x<double>, const tensor<double> & y) {
  // This function written for homogeneous types, but not unnecessary
  double tot = 0.0;
  for_each_tensors([&tot](double xV, double yV) {
    tot += xV * yV;
  },
  x.data_shape(), /* Iterate over valid tuples for x.data_shape(); as written, this line assumes x has smaller
      ↪ shape */
  x, y);
  return tot;
}
```

function ( **Algorithm 9**). Likewise, this pattern can be trivially extended to pass a const view of the tuple counter through as the first argument to function (**Algorithm 10**). When combined with variable capture, this enables fast vector operations with a much greater generality. For example, numpy is very efficient for straightforward vector operations, but loops must be used when the tuple index is needed. For example, there is no native way in numpy to efficiently find the bounding box containing all nonzero values in a tensor. Likewise, the type of iteration used by libraries like boost [2] to vectorize operations over tensors is inherently hierarchical, and thus does not have access to the index tuple. Using the template-recursive iteration over tensors (TRIOT) pattern, it would be trivial to find the nonzero bounding box for the element-wise product of two tensors. Importantly, classic and numerically demanding problems such as marginalization and expected value computation require the tuple index to compute their results (simple example shown in **Algorithm 11**).

■ **Algorithm 9** Alternative functions to allow external modifications of the tensors. modify_tensors allows modification of all arguments, apply_tensors only allows modification to the destination argument. Regardless, the ForEachFixedDimension class is used and the appropriate const type is automatically detected by the compiler.

```
 1  // Allows modifications to all arguments:
 2  template <typename FUNCTION, typename ...DEST_TENSORS>
 3  void modify_tensors(FUNCTION function, const Vector<unsigned long> & shape, DEST_TENSORS && ...args)
    ↪ {
 4    check_tensor_pack_bounds<DEST_TENSORS...>(args..., shape);
 5    LinearTemplateSearch<0u,MAX_TENSOR_DIMENSION,TRIOT::ForEachFixedDimension>::apply(shape.size(),
    ↪ shape, function, args...);
 6  }
 7
 8  // Allows modifications only to dest:
 9  template <typename FUNCTION, typename DEST_TENSOR, typename ...SOURCE_TENSORS>
10  void apply_tensors(FUNCTION function, const Vector<unsigned long> & shape, DEST_TENSOR && dest,
    ↪ const SOURCE_TENSORS & ...source_args) {
11    check_tensor_pack_bounds<DEST_TENSOR, SOURCE_TENSORS...>(dest, source_args..., shape);
12    LinearTemplateSearch<0u,MAX_TENSOR_DIMENSION,TRIOT::ForEachFixedDimension>::apply(shape.size(),
    ↪ shape, function, dest, source_args...);
13  }
```

## 3 Results

The TRIOT pattern is compared to alternative methods: tuple iteration; tuple iteration where the dimension is known at compile time; integer reindexing; integer reindexing where the axes are restricted to powers of 2 (see **Algorithm 12**; this is an unrealistic constraint in practice, but it is interesting to investigate the performance gains from converting % operations to & operations and / and * operations to » and « operations, respectively); numpy; boost::multi_array (using version 1.61.0); C-style for loops (hard coded for each problem size); vectorized Fortran code; for loops in Go. Note that of these methods only tuple iteration, integer reindexing, numpy, and TRIOT are possible when the dimension is not known at compile time.

These methods are compared on four test problems: In benchmark 1, data is copied from a tensor of shape $(2^{10}, 2^9, 2^8)$ to a tensor of shape $(2^9, 2^9, 2^5)$. In benchmark 2, an inner product between two tensors of shape $(2^{10}, 2^9, 2^8)$ and $(2^9, 2^9, 2^5)$ is computed (visiting only tuple indices shared by both). In benchmark 3, $x[t] \leftarrow x[t] + y[t] \cdot x[t] - z[t]$ is performed for all tuples $t$ that are valid with respect to the shape of $x$, using $x.data\_shape() = (2^7+1, 2^5, 2^3+5, 2^4)$, $y.data\_shape() = (2^8-3, 2^6, 2^6, 2^4+7)$, and $z.data\_shape() = (2^8, 2^5+7, 2^6, 2^5+1)$. In benchmark 4, implementations of multidimensional convolution via tuple iteration (**Algorithm 14**), TRIOT (**Algorithm 15**), and numpy are compared by convolving two matrices, each with shape $(2^8, 2^3)$. Although in scientific computing the fast Fourier transform (FFT) convolution is often used for large convolutions, small problems are often much faster to solve using the naive approach demonstrated here.

Wherever possible, the competing methods (but not TRIOT) are allowed to exploit the fact that lexicographically iterating over all tuples in tensor $x$ is the same as

**Algorithm 10** Providing safe access to the current tuple index. As the loops iterate, a const view of the counter will be sent as the first argument of function.

```cpp
template <unsigned int DIMENSION, unsigned int CURRENT>
class ForEachVisibleCounterFixedDimensionHelper {
public:
  template <typename FUNCTION, typename ...TENSORS>
  inline static void apply(tup_t counter, const_tup_t shape, FUNCTION function, TENSORS & ...args) {
     for (counter[CURRENT]=0; counter[CURRENT]<shape[CURRENT]; ++counter[CURRENT])
        ForEachVisibleCounterFixedDimensionHelper<DIMENSION—1, CURRENT+1>::template apply<
            ↪ FUNCTION, TENSORS...>(counter, shape, function, args...);
  }
};

template <unsigned int CURRENT>
class ForEachVisibleCounterFixedDimensionHelper<1u, CURRENT> {
public:
  template <typename FUNCTION, typename ...TENSORS>
  inline static void apply(tup_t counter, const_tup_t shape, FUNCTION function, TENSORS & ...args) {
     for (counter[CURRENT]=0; counter[CURRENT]<shape[CURRENT]; ++counter[CURRENT])
     // Cast the counter to a const_tup_t pointer so that its
     // contents cannot be modified by function:
     function(static_cast<const_tup_t>(counter), CURRENT+1, args[tuple_to_index_fixed_dimension<
         ↪ CURRENT+1>(counter, args.data_shape())]...);
  }
};

template <unsigned char DIMENSION>
class ForEachVisibleCounterFixedDimension {
public:
  template <typename FUNCTION, typename ...TENSORS>
  inline static void apply(const_tup_t shape, FUNCTION function, TENSORS & ...args) {
      unsigned long counter[DIMENSION];
      memset(counter, 0, DIMENSION*sizeof(unsigned long));
      ForEachVisibleCounterFixedDimensionHelper<DIMENSION,0>::template apply<FUNCTION, TENSORS...>(
          ↪ counter, shape, function, args...);
    }
  };

template <typename FUNCTION, typename ...TENSORS>
void enumerate_for_each_tensors(FUNCTION function, const Vector<unsigned long> & shape, const
    ↪ TENSORS & ...args) {
  check_tensor_pack_bounds<TENSORS...>(args..., shape);
  LinearTemplateSearch<0u,MAX_TENSOR_DIMENSION,TRIOT::ForEachVisibleCounterFixedDimension>::
      ↪ apply(shape.size(), shape, function, args...);
}
```

# TRIOT: Faster tensor manipulation in C++11

■ **Algorithm 11** Example of TRIOT using the index tuple. The index tuple is passed to the function to compute a sum over all tuples, as weighted by all tensor arguments.

```
1  Vector<double> result(DIMENSION);
2  result.fill(0.0);
3  enumerate_for_each_tensors([&result](const_tup_t counter, const unsigned int dim, double xV, double yV,
       ↪ double zV) {
4    /* Could alternatively capture the variable DIMENSION instead of passing in unsigned int dim; however,
       ↪ as written, even though dim is not constexpr, it will essentially become a compile—time
       ↪ constant due to constant propagation. This can permit greater optimization. */
5    for (unsigned int k=0; k<dim; ++k)
6      result[k] += counter[k] * xV * yV * zV;
7  },
8  x.data_shape(),
9  x, y, z);
```

■ **Algorithm 12** Reindex flat index from one tensor shape to another tensor shape (special case where axes are all powers of 2). The algorithm takes a flat integer index with respect to log_shape and converts it to a flat integer index with respect to new_log_shape, where both shapes have length dimension.

```
1  inline unsigned long reindex_powers_of_2(unsigned long index, const unsigned int*__restrict const
       ↪ log_shape, const unsigned int*__restrict const new_log_shape, unsigned int dimension) {
2    unsigned long new_index = 0;
3    unsigned int new_axis_sum_from_right = 0;
4    for (int i=dimension—1; index>0 && i>=0; ——i) {
5      unsigned long next_log_axis = log_shape[i];
6      unsigned long new_next_log_axis = new_log_shape[i];
7
8      unsigned long next_value = index & ((1ul<<next_log_axis)—1);
9
10     new_index += (next_value << new_axis_sum_from_right);
11     index >>= next_log_axis;
12
13     new_axis_sum_from_right += new_next_log_axis;
14   }
15   return new_index;
16 }
```

iterating over all flat indices in order. The reported runtimes measure only the desired task; the cost of initialization is not included. On every benchmark, each method was run 128 times on a cold boot of Ubuntu 16.04 with no other programs running and the X server shut down (via sudo service lightdm stop). Runtimes are reported in **Figures 1–4**. These benchmarks were run on an 2.0 GHz Intel Core i7 with 256KB of L1 cache, 1MB of L2 cache, 6MB of L3 cache, and 8GB RAM. All C / C++ programs (including boost) were compiled with clang++ (version 3.8), g++ (version 5.4) and Intel icc (version 17.0 using trial license) with the following flags: -std=c++11 -Ofast -march=native -mtune=native -fomit-frame-pointer. The Fortran programs were written using Fortran 95 and compiled with f95 -Ofast -march=native -mtune=native -fomit-frame-pointer. All Fortran implementations use axes in reversed order and access data in a cache-optimized fashion, because Fortran uses column-major array format. All methods also had bounds checking explicitly disabled for greater speed (*e.g.*, using #define NDEBUG and #define BOOST_DISABLE_ASSERTS for the boost benchmarks). Benchmark 1 was the only benchmark to which all methods could be applied; therefore, the respective implementations of each method for benchmark 1 are shown in **Algorithm 13**.

The principal motivation with all four benchmarks is to address accessing data in tensors of different shapes (which will not be accessed in a contiguous fashion). Benchmark 1 is reminiscent of extracting a smaller tensor slice that is embedded in a larger tensor; this is an operation commonly performed when when extracting an FFT convolution result from the zero-padded tensor used for FFT.

All benchmarks 1–4 are challenged by data movement and memory access patterns. Benchmarks 2–4 also include floating point arithmetic. In benchmarks 1, 3 and 4 data is directly written to a tensor; this challenges attention to pointer aliasing, which may cause slower performance when the method does not provide explicit guarantees that the source and destination tensors refer to different regions of memory.

boost was not applicable to benchmarks 2 and 3 because no operators are provided for boost::multi_array, and performing the inner product on tensors of different shapes (benchmark 2), i.e. on non-overlapping flat indices, is difficult to do via iterators (because the shapes iterated over do not match); however, benchmark 2 is designed with the same number of tensors and same shape as benchmark 1, in which boost was comparable to tuple iteration. As such, benchmark 2 and benchmark 3 would essentially need to be written using nested for loops in a manner identical to the C method. Benchmark 2 is reminiscent of a marginalization of a product between two distributions and benchmark 3 resembles the calculation of a covariance. Benchmark 4 challenges the flexibility to efficiently manipulate and use an index tuple on the fly.

■ **Algorithm 13** Various implementations of benchmark 1.

```
1  // Tuple iteration (DIMENSION must be compile—time constant):
2  vector<unsigned long> t(DIMENSION);
3  t.fill(0);
4  unsigned long k;
5  for (k=0; k<x.flat.size(); advance_tuple_fixed_dimension<DIMENSION>(&t[0], &x.data_shape()[0]), ++k)
6    x[k] = y[tuple_to_index_fixed_dimension<DIMENSION>(&t[0], &y.data_shape()[0])];
7
8  // boost:
9  x[ boost::indices[range(0, x.shape[0])][range(0,x.shape[1])][range(0,x.shape[2])] ] = y[ boost::indices[range(0,
       ↪ x.shape[0])][range(0,x.shape[1])][range(0,x.shape[2])] ];
```

## TRIOT: Faster tensor manipulation in C++11

```
! Fortran 95
! Note that the slice is taken with the axis order reversed to guarantee greater cache performance,
↪ because Fortran stores data in column-major format.
x = y(1:2**5,1:2**9,1:2**9)

// Hard-coded for loops in C:
unsigned long k;
for (k=0; k<x.data_shape()[0]; ++k) {
    for (unsigned long j=0; j<x.data_shape()[1]; ++j) {
        unsigned long x_bias = (k*x.data_shape()[1] + j)*x.data_shape()[2];
        unsigned long y_bias = (k*y.data_shape()[1] + j)*y.data_shape()[2];
        for (unsigned long i=0; i<x.data_shape()[2]; ++i)
            x[x_bias + i] = y[y_bias + i];
    }
}

// Integer reindexing:
unsigned long k;
for (k=0; k<x.flat.size(); ++k)
  x[k] = y[reindex(k, &x.data_shape()[0], &y.data_shape()[0], DIMENSION)];

// Integer reindexing (axes are powers of 2):
unsigned long k;
for (k=0; k<x.flat.size(); ++k)
  x[k] = y[reindex_powers_of_2(k, &x_log_shape[0], &y_log_shape[0], DIMENSION)];

// Tuple iteration (DIMENSION unknown at compile time):
vector<unsigned long> t(DIMENSION);
t.fill(0);
unsigned long k;
for (k=0; k<x.flat_size(); advance_tuple(&t[0], &x.data_shape()[0], DIMENSION), ++k)
  x[k] = y[t];

# numpy (python):
# Note that this is hard coded for DIMENSION=3, but it can be implemented generically via list
↪ comprehensions; however, this is hard-coded so as to not penalize numpy for python's
↪ interpreter and let the code get to numpy's underlying C/Fortran underpinnings as quickly as
↪ possible. Also note that the slice of y must be cast as a new array in order to make a copy (rather
↪ than simply point x at a reference slice of y).
x_sh = x.shape.
x = np.array(y[:x_sh[0], :x_sh[1], :x_sh[2]])

// Go:
for i := 0; i < 1<<9; i++ {
  for j := 0; j < 1<<9; j++ {
    for k := 0; k < 1<<5; k++ {
      x[i][j][k] = y[i][j][k]
    }
  }
}

// TRIOT (DIMENSION unknown at compile time):
apply_tensors([](double & xV, double yV) {
  xV = yV;
  },
  x.data_shape(),
  x, y);
```

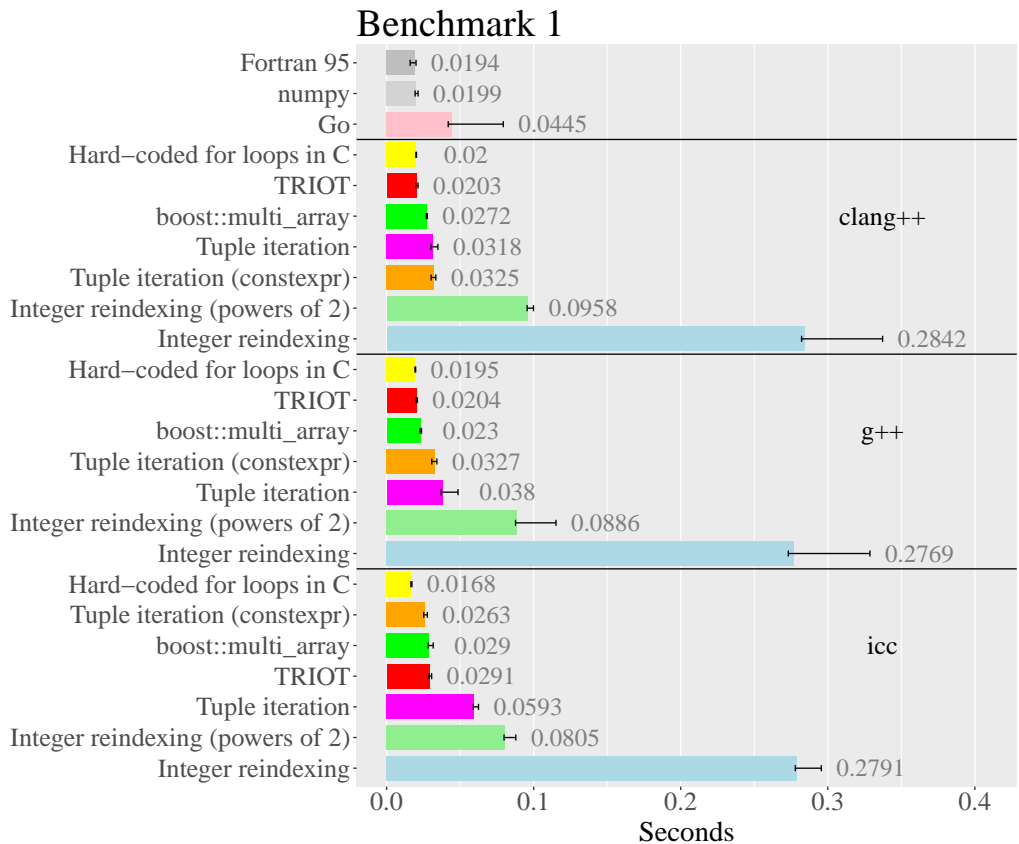

**Figure 1** **Benchmark 1: Runtimes (in seconds) of copying a tensor into a tensor of different shape.** The error bars represent the minimum and maximum runtimes measured for each method and the gray number the mean runtime. Integer reindexing is the slowest; even when the axes are restricted to powers of 2 and bit twiddling is used for division, multiplication, and modulo operation, reindexing is slower than tuple iteration, which does not bode well for ideas that would attempt to embed tensors into larger shapes with relatively prime indices and perform fast reindexing via the Chinese remainder theorem. Tuple iteration performed better than Go and substantially better than reindexing, but is slower than boost. boost is slower than vectorized Fortran, numpy, hard-coded C-style looping, and TRIOT, which all produce similar runtimes, regardless of whether clang++ or g++ is used. With the exception of simple nested for loops in C, icc does not optimize quite as effectively on this benchmark. Of the fastest methods, only numpy and TRIOT do not need to know the dimension in advance.

**TRIOT: Faster tensor manipulation in C++11**

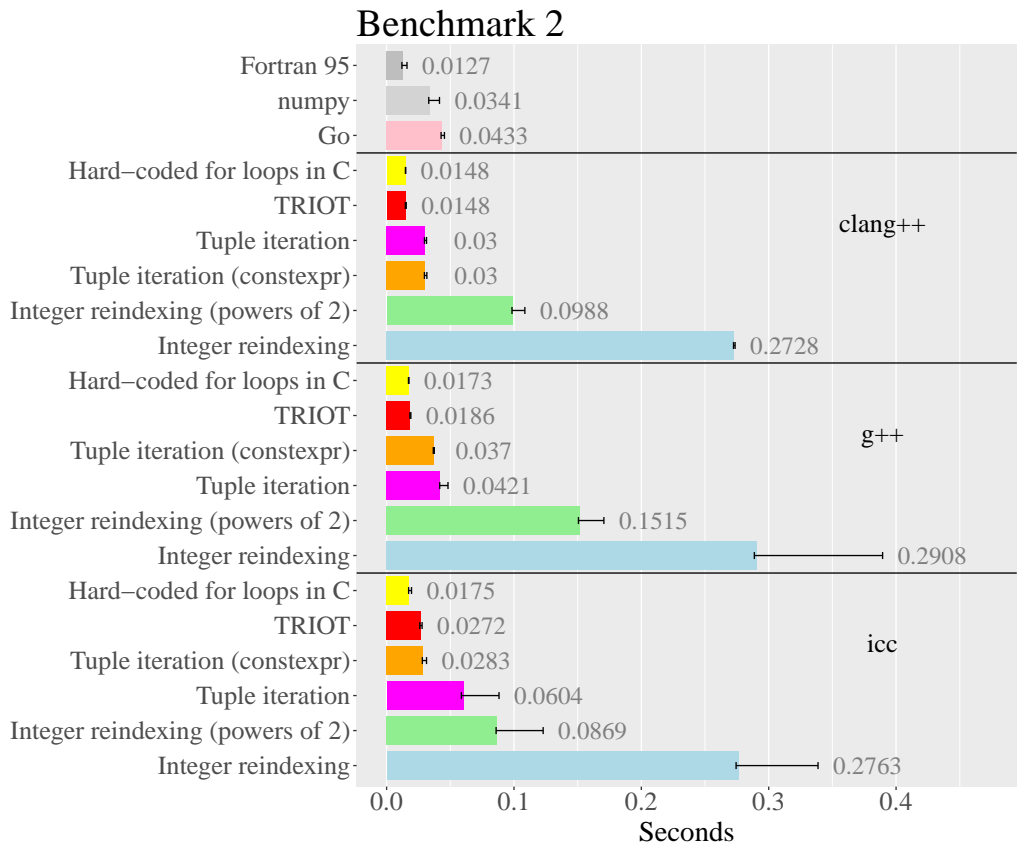

**Figure 2 Benchmark 2: Runtimes of inner product between tensors (in seconds) for all compilers.** The error bars represent the minimum and maximum runtimes measured for each method and the gray number the mean runtime. Integer reindexing is the slowest and the vectorized Fortran the fastest method; Fortran is similar in performance to TRIOT and hard-coded for loops in C.

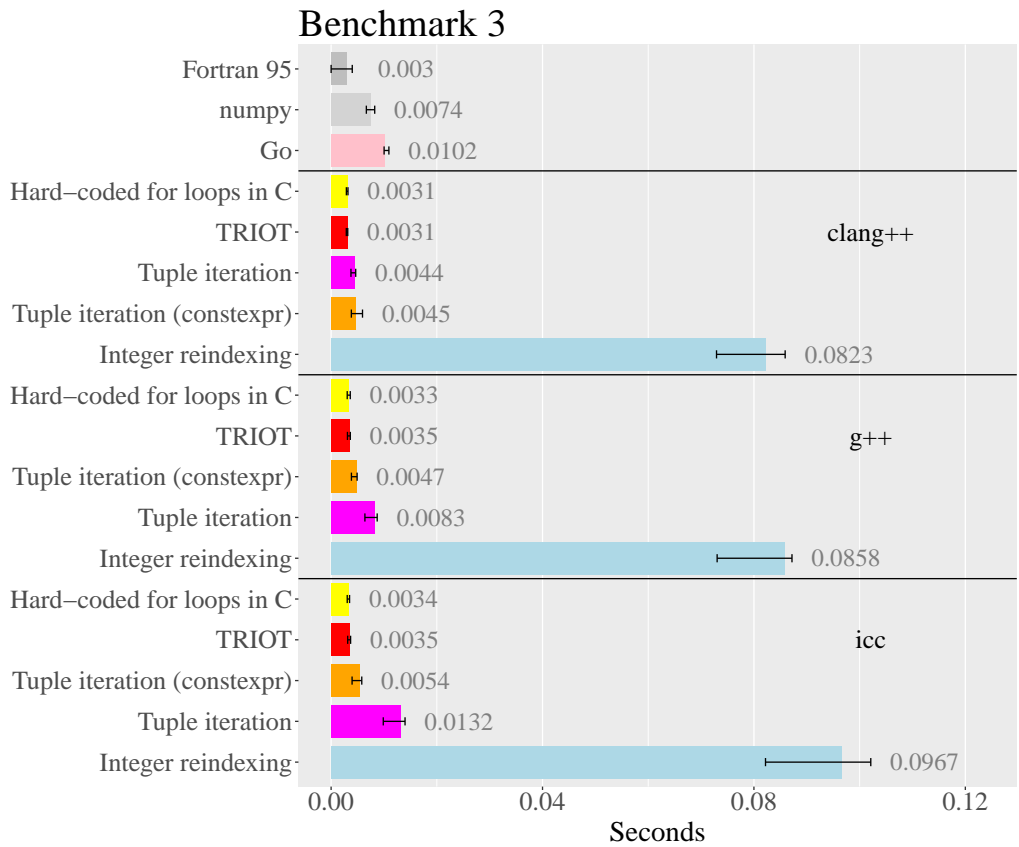

**Figure 3** Benchmark 3: Runtimes of operations broadcast over tensors (in seconds) for all compilers. The error bars represent the minimum and maximum runtimes measured for each method and the gray number the mean runtime. Integer reindexing is the slowest followed by a faster tuple iteration and the implementation in numpy. The performance of TRIOT, hard-coded for loops in C, and vectorized Fortran are similar to one another and are faster than the other methods, regardless of the compiler used. Note that the large error bars from Fortran are partly due to low precision of the dtime function, which sometimes suffers underflow for small runtimes.

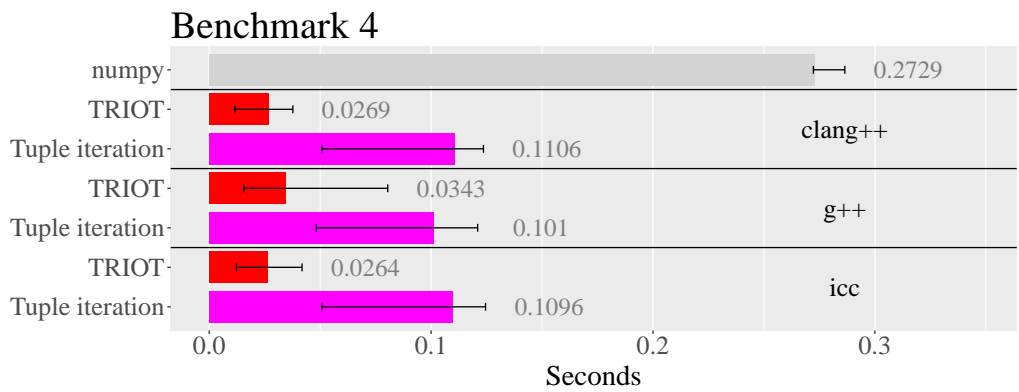

**Figure 4** **Benchmark 4: Runtimes (in seconds) of multidimensional naive convolution.** The error bars represent the minimum and maximum runtime of each method and the gray number the mean runtime. Convolving two matrices with shapes $(2^8, 2^3)$ with TRIOT is $> 3\times$ faster than with tuple iteration and $> 7\times$ faster than the scipy.signal.convolve function.

- **Algorithm 14** Multidimensional convolution via tuple iteration.

```
1  Tensor<double> tuple_iteration_naive_convolve(const Tensor<double> & lhs, const Tensor<double> & rhs)
       ↪ {
2    assert(lhs.dimension() == rhs.dimension());
3
4    Tensor<double> result(lhs.data_shape() + rhs.data_shape() − 1ul);
5    result.flat().fill(0.0);
6    Vector<unsigned long> counter_lhs(lhs.dimension());
7    Vector<unsigned long> counter_rhs(rhs.dimension());
8    Vector<unsigned long> counter_result(result.dimension());
9
10   for (unsigned long i=0; i<lhs.flat_size(); ++i, advance_tuple(counter_lhs, lhs.data_shape(), lhs.dimension()
       ↪ )) {
11     counter_rhs.fill(0ul);
12     for (unsigned long j=0; j<rhs.flat_size(); ++j, advance_tuple(counter_rhs, rhs.data_shape(), rhs.
         ↪ dimension())) {
13       for (unsigned int i=0; i<result.dimension(); ++i)
14         counter_result[i] = counter_lhs[i] + counter_rhs[i];
15
16       unsigned long lhs_flat = tuple_to_index(counter_lhs, lhs.data_shape(), lhs.dimension());
17       unsigned long rhs_flat = tuple_to_index(counter_rhs, rhs.data_shape(), rhs.dimension());
18       unsigned long result_flat = tuple_to_index(counter_result, result.data_shape(), result.dimension());
19       result.flat()[result_flat] += lhs.flat()[lhs_flat] * rhs.flat()[rhs_flat];
20     }
21   }
22
23   return result;
24 }
```

## 4 Discussion

In the benchmarks performed, clang++ was shown to be the fastest compiler for TRIOT, perhaps due to clang++'s advanced support of optimizations based on constant propagation, which pair well with the many constexpr values used in the TRIOT implementation. In spite of the greater flexibility of TRIOT operations and the fact that tensor dimensions need not be known at compile time, TRIOT is consistently competitive with C and Fortran and frequently outperforms the other methods tested.

Although this implementation relies on $\lambda$-functions, the use of $\lambda$-functions is not necessary for the general method. The $\lambda$-function could possibly be replaced by expression templates [5, 6], which could be generated in the standard manner, via the parse trees produced automatically at compile time by defining operators on tensors (+, -, *etc.*); however, the generality permitted by variable capture and by directly accessing the index tuple is a great benefit to the $\lambda$-function implementation. Furthermore, unlike expression templates, no care needs to be taken to prevent parse trees from obscuring underlying mathematical properties (*e.g.*, x[t] = y[t] + y[t] + y[t] + $\cdots$ can sometimes be optimized as x[t] = y[t]*K for some unsigned integer K known at compile time). Because the function code is never converted into a data structure and then back into code, such considerations are unnecessary.

■ **Algorithm 15**    Multidimensional convolution via TRIOT.

```
1  Tensor<double> triot_naive_convolve(const Tensor<double> & lhs, const Tensor<double> & rhs) {
2    assert(lhs.dimension() == rhs.dimension());
3
4    Tensor<double> result(lhs.data_shape() + rhs.data_shape() — 1ul);
5    result.flat().fill(0.0);
6    Vector<unsigned long> counter_result(result.dimension());
7
8    enumerate_for_each_tensors([&counter_result, &result, &rhs](const_tup_t counter_lhs, const unsigned
         ↪ int dim_lhs, double lhs_val) {
9      enumerate_for_each_tensors([&counter_result, &result, &rhs, &counter_lhs, &lhs_val](const_tup_t
           ↪ counter_rhs, const unsigned int dim_rhs, double rhs_val) {
10       for (unsigned int i=0; i<dim_rhs; ++i)
11         counter_result[i] = counter_lhs[i] + counter_rhs[i];
12       unsigned long result_flat = tuple_to_index(counter_result, result.data_shape(), dim_rhs);
13       result.flat()[result_flat] += lhs_val * rhs_val;
14     },
15     rhs.data_shape(),
16     rhs);
17   },
18   lhs.data_shape(),
19   lhs);
20
21   return result;
22 }
```

Also, because of the way that operations are vectorized, it is easy to safely read and write from the same tensor (*e.g.*, x[t] += y[t] * x[t]) by simply applying a two-argument $\lambda$-function with a reference parameter (for the x[t] reference) and one value parameter (for the y[t] value). apply_tensors would be called with only two tensor arguments x, y. Pairing TRIOT with expression templates would make this guarantee much more difficult, since the tensor x would appear multiple times in the expression tree. This distinction is important to allow the underlying flat vector pointer to be of type T*__restrict, which guarantees the compiler that it will not be modified from one view while read from another view. The end result is that values indexed by such pointers do not need to be re-read from memory when another pointer is dereferenced and changed. This "pointer aliasing" problem and the making of copies when using operators (+, -, …) on tensors are together often referred to casually as the only reasons that Fortran is faster than C / C++, but __restrict pointers would not be an issue with TRIOT, because the value or reference can occur multiple times in the $\lambda$-function, but the tensor reference need only be included once. The runtimes presented use __restrict pointers in the vector<T> type, and so the hard-coded for loops in C, vectorized Fortran, and TRIOT methods are all comparable.

Bounds checking can also be performed to make code safe but perform a single check over shapes rather than over indices (this would make a significant difference compared to boost, where bounds checking on the benchmark problems yielded 2× the runtime of boost without bounds checking). The method can also be trivially paired with slices, which reference the original tensor and store a bias added when performing all flat index operations. For example, for a tensor $x$ with shape $x.data\_shape()$ and dimension $d = x.dimension()$, a tensor slice starting at index tuple $(i_1, i_2, \ldots, i_d)$ can be constructed by using a bias: $bias \leftarrow \texttt{tuple\_to\_index}((i_1, i_2, \ldots, i_d), x.data\_shape(), d)$. Any valid accesses to the tensor slice $x\_slice[flat\_index]$ can be performed as $x.flat[bias + flat\_index]$.

Parallelism could be automatically exploited by specializing the classes in the template recursion that correspond to the outermost for loops (template <unsigned int DIMENSION> class ForEachVisibleCounterFixedDimensionHelper<DIMENSION, 0>, template <unsigned int DIMENSION> class ForEachVisibleCounterFixedDimensionHelper<DIMENSION, 1>, *etc.*). This would allow multiple processors, cores, or GPUs to simultaneously process partitions of tensors. The limitation of GPUs is frequently the cost of data movement: the processing bandwidth of a GPU often far exceeds the bandwidth of the bus. For this reason, a GPU speed-up is most likely when operating on a large, contiguous block of memory, which is allocated on the GPU or copied to the GPU in chunks, which are subsequently processed on the GPU for a long duration. Since operations on multiple tensors of different shapes will access non-contiguous blocks of memory, a GPU-based TRIOT implementation may still be much slower than the CPU-based TRIOT implementation on some problems. Specifically, when the innermost for loop iterates over several indices and when the numeric operations applied to each element require several steps each (improving the benefit of parallelization), then a GPU-based implementation may be worthwhile. Even when GPU-based TRIOT implementation is particularly efficient due to the small contiguous blocks of memory accessed by the innermost for loop (*e.g.*, when operating on only the $(n, 4)$ submatrix of a matrix with shape $(n, n)$), TRIOT can be used to efficiently move the relevant portion of the tensor into a contiguous block of memory allocated via cudaMalloc, which can then be processed by the GPU. Regardless, using multiple cores to process a TRIOT expression can be used to parallelize any arbitrary for loop (not necessarily the innermost one) and will likely provide a reasonable speedup in the general case because CPUs commonly offer a dedicated cache for each core. Multiple parallel threads that modify the same variable (via variable capture in the lambda functions) may result in undefined behavior, and so applications where it's essential (*e.g.*, computing an inner product) could rely on a reduce-like methodology, in which each thread computes its own pre-result and these pre-results are amalgamated together to produce the final result.

The only real downside of the proposed method is compile time, which increases with the maximum dimension permitted: When the maximum dimension is set to 16, it compiles in 16 seconds and when the maximum tensor dimension is set to 32, it compiles in 1 minute; however, this can be solved easily by compiling the tensor code offline so that the resulting library is already optimized for all realistic tensor dimensions. Note that if no axes are trivial, then the memory usage to store the tensor must grow in $\Omega(2^d)$, so it is reasonable to assume that the dimensionality should be limited; if some axes have trivial length 1, then those axes can simply be ignored to operate on a tensor in a lower dimension. Even if the dimensionality is not known at compile time, a tight bound on the maximum dimensionality should be reasonably known (in comparison, Fortran 95 only supports dimensions up to 7). Furthermore, the compile time is reduced when dimensionality of TRIOT expressions can be inferred at compile time (the $d_{\max} = 32$ case compiled in 3 seconds).

Importantly, TRIOT expressions use universal references in C++11 (that is, template types paired with rvalue references). This permits any type to be used in TRIOT expressions: value parameters, rvalue references, references, and const references. The practical implication of this is that tensor views, non-contiguous regions of a particular window of a tensor, can be sent into TRIOT expressions without needing to declare a separate variable before the TRIOT expression (as value parameters, they will be interpreted by the compiler as rvalue references). This feature is useful if you want to compute, for example, the product of two or more probability mass functions with varying support: Tensor views can first be constructed, which window smaller regions of the tensors over their intersecting support. Then, these tensor views are operated on by TRIOT expressions as if they were themselves tensors. This allows TRIOT expressions to be used without allocating memory for these views (they simply refer to the memory in the original tensor arguments).

The real advantage of this pattern is in its simultaneous simplicity and flexibility, which allows high performance while still allowing both the dimension and the shape to be unknown at compile time with a simple interface to the programmer. This is a large advantage over libraries like boost, which requires more complicated code, is slower in benchmarks, requires the dimension of all tensors must be known at compile time, and does not allow access to the tuple index inside a vector operation. More flexible methods for easy, efficient manipulation of tensors are crucially important for computing in fields of research where the dimensionality may be changing dynamically during runtime, for example iterative machine learning algorithms for which the sparsity of solutions increases as convergence is reached, thus restricting tensors to subspaces of lower dimension.

## 5 Availability

A small tensor library demonstrating these ideas (including demonstrations of how TRIOT would be used for a few problems and how tensor views are used) is available at https://bitbucket.org/orserang/triot. It is licensed with a Creative Commons Attribution 4.0 International license, and can be freely used in both academic or commercial software.

**Acknowledgements** We are grateful to Seth Just for his helpful comments and to Mattias Frånberg for the discussion.

## About the authors

**Florian Heyl** is completing his master Praktikum in the Institute of Computer Science at Freie Universität Berlin.

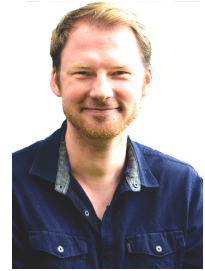

**Oliver Serang** is a professor in the Institute of Computer Science at Freie Universität Berlin. He can be reached at orserang@uw.edu.

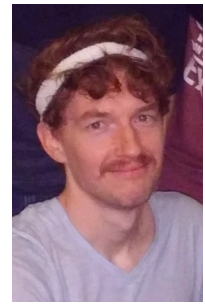